\documentclass[runningheads]{llncs}
\usepackage{graphicx}
\usepackage{amsmath}
\usepackage{algpseudocode}
\usepackage[ruled]{algorithm2e} % For 

\SetCommentSty{mycommfont}
\usepackage[caption=false]{subfig}
\usepackage{amssymb}
\usepackage{color}
\usepackage{array}
\usepackage{multirow}
\usepackage{float}
\begin{document}
%
%\title{Using network motifs as a generative tool for understanding information flow in cascades}
\title{Understanding Information Flow in Cascades Using Network Motifs}
\author{Soumajyoti Sarkar \and
Hamidreza Alvari  \and
Paulo Shakarian}
\authorrunning{Sarkar et al.}
\institute{Arizona State University, Tempe, Arizona, USA \and
\email{\{ssarka18, halvari, shak\}@asu.edu}}
\maketitle              % typeset the header of the contribution
\begin{abstract}
A growing set of applications consider the process of network formation by using subgraphs as a tool for generating the network topology. One of the pressing research challenges is thus to be able to use these subgraphs to understand the network topology of information cascades which ultimately paves the way to theorize about how information spreads over time. In this paper, we make the first attempt at using network motifs to understand  whether or not they can be used as generative elements for the diffusion network organization during different phases of the cascade lifecycle. In doing so, we propose a motif percolation-based algorithm that uses network motifs to measure the extent to which they can represent the temporal cascade network organization. We compare two phases of the cascade lifecycle from the perspective of diffusion-- the phase of steep growth and the phase of inhibition prior to its saturation. Our experiments on a set of cascades from the Weibo platform and with 5-node motifs demonstrate that there are only a few specific motif patterns with triads that are able to characterize the spreading process and hence the network organization during the inhibition region better than during the phase of high growth. In contrast, we do not find compelling results for the phase of steep growth.

\end{abstract}
\section{Introduction}
Structural patterns within networks have been widely used to understand the mechanisms of information diffusion, especially in the contexts of information cascades~\cite{jure_2007} and social networks \cite{Backstrom_2006} where %the authors study the 
community membership is studied as a measure of the growth of cascades. %Extending these, 
There exist several studies \cite{Ashton_2015,Goel_2012} that have attempted at understanding the structural virality of the cascade network structures to understand why some cascades experience a large growth in adoption%and why some cascades would not experience much popularity
, by either confirming or countering the hypothesis of tightly-knit communities or clusters being obstructions to information flow. Inspired by these prior studies, we make the first attempt at understanding the network organization of information cascades in two stages of the cascade lifecycle: (1) the phase where a cascade experiences a sudden momentum in adoption (steep growth) and (2) the phase where the cascade approaches inhibition prior to complete stagnation (inhibition). 

To this end, we use network motifs patterns \cite{Milo} of varying densities as basic starting elements and devise a percolation algorithm to measure the extent to which we could replicate the network organization of the cascades in these phases using these motifs. This helps us investigate the two important objectives: (1) it allows us to understand the network formation process in these phases as opposed to analyzing the structural virality or patterns over the entire cascade as done before, and (2) it allows us to now compare these two phases by scrutinizing which patterns distinguish these phases in terms of whether specific patterns act as better generative elements in one of these phases. Overall, it assists in unveiling the complex network formation process during information diffusion that cannot be simply done by measuring the graph densities, the clustering coefficient or other network metrics on the cascade as a whole.

One factor that distinguishes our work from others is that we do not dissect the social network or any time stamped historical network to analyze what patterns are significant from the perspective of their frequencies \cite{Paranjape}-- this helps in measuring the network structure merely from the perspective of communities and the clusters and how they could be used to understand information passage or blockage. Rather, our motivation in using these network motifs is to see whether or not they can be used as basic building blocks to replicate the information spreading process-- this cannot be achieved using disjoint instances of these motifs and counting their frequency of appearances. In other words, we start with motif patterns and try to build up the network structure of the cascades, much like a social contagion process except we use motif blocks instead of individual nodes. More specifically, we would like to know if denser patterns are more successful in replicating the inhibition phase network compared to the steep phase. It could explain why information blocks in inhibition or the presence of linear chain and star shapes in the steep phase represents structural virality.  Our main contributions in this paper are as follows:
\vspace*{-2mm}
\begin{enumerate}
    \item We use 5-node network motifs to compare the structural organization of the diffusion networks during two periods of the cascade lifecycle : the period of steep growth and the period of inhibition.
    \item We propose a new motif percolation-based algorithm that reconstructs the temporal diffusion network structure in these two periods by cascading the motifs across neighborhoods in these two periods.
    \item Our results on 7407 cascades from Weibo suggest that while the acyclic patterns such as \ \includegraphics[scale=0.06]{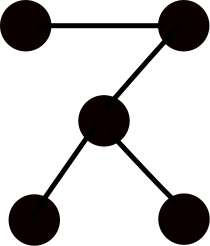}, \ \includegraphics[scale=0.06]{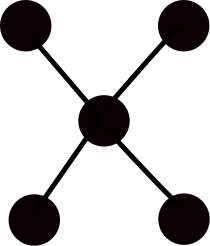} \ or very dense patterns such as \ \includegraphics[scale=0.06]{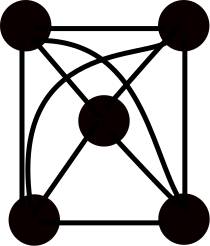} \  and \ \includegraphics[scale=0.06]{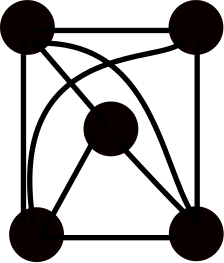} \ are not able to distinguish these two phases, it is mainly the patterns containing both triads and linear chains \ \includegraphics[scale=0.06]{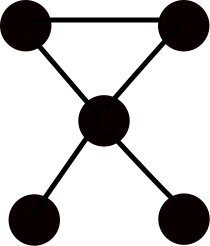}, \ \includegraphics[scale=0.06]{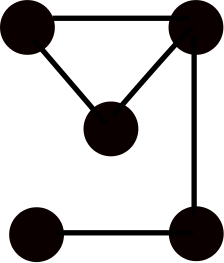} that explain the difference in these two phases - the inhibition phase being characterized by these generative motifs better than the steep phase.
\end{enumerate}
\vspace*{-4mm}
\section{ Information cascades}
\label{sec:tech_prelim}
Formally, a cascade $C$ can be represented by a sequence of $3$-tuples denoted by $(u, v, t) $ which indicates that the microblog was reshared by $v$ from the user $u$ at time $t$. We denote the sequence of reshare times for $C$ as $\tau_C$ = $\langle 0, \dots, t, \ldots T_C \rangle$ ordered by time, $T_C$ being the time difference between the first and last posting. We will often use the notation $\tau'$ to denote a subsequence of $\tau$. In our work, we create the sequence of subsequences $\tau$ = $\langle \tau'_1, \ldots, \tau'_{\mathcal{Q}} \rangle$, ordered by the starting time of each subsequence, where we denote $ \mathcal{Q}$ to be the number of subsequences for $C$, which would vary for different cascades. The method of splitting the cascade $C$ into a sequence of such subsequences $\tau'$ has been described in detail in Section~\ref{sec:dataset}. 

Rules for identifying cascade stages or subsequences mapping the intervals of maximum growth which we call henceforth the ``steep" interval and the ``inhibition" interval are generally not well defined in the context of information cascades. In order to be able to compare the diffusion processes happening at these two stages of the cascade lifecycle, the first problem we address is to identify 2 subsequences: $\tau'_{steep}$ during which the cascade experiences the maximum rate in reshares and $\tau'_{inhib}$ during which the cascade fails to regain any momentum in growth and finally expires. We adopt a 3-step procedure\footnote{https://bit.ly/2tN4fnQ} to detect these 2 intervals as done in \cite{sarkar} and we use them to compare the diffusion process from the perspective of topology and network motifs. At the end of the procedure, we obtain the time points $t_{steep}$ and $t_{inhib}$ $\in$ $[0, T_C]$, and refer to the subsequences $\tau' \in \tau$ containing these time points as $\tau'_{steep}$ and $\tau'_{inhib}$ respectively. 

\iffalse
\subsection{Steep and Inhibition Intervals}
\label{sec:steep_inhib}
We identify these intervals in a three-step process, which we provide technical details for in the Appendix A1. This process is described intuitively below:
 \begin{enumerate}
 	\item We calculate the Hawkes intensity \cite{seismic}\cite{hawkes_zha}  at each reshare time point $t$ as a function of the number of past interactions, and the distribution of times taken by the users to adopt the cascade $C$.
 	\item\label{step2} We then identify intervals with local maxima (which are candidates for the steep interval) and local minima (which are candidates for the inhibition interval).
 	\item  We then use a maximum-likelihood approach to filter the points in step~\ref{step2} and obtain the parameters that we use to infer the \textit{steep} and \textit{inhibition} times of the new cascades.
 \end{enumerate}
 
 \noindent Once we infer the parameters, we follow the above procedure for identifying the \textit{event times} (which we represent by the mean of the respective event intervals) of the rest of the cascades. While doing so, we ignore the maximum likelihood step and directly use the inferred parameters to compute the event times using a threshold technique. 
 \fi

\vspace*{-4mm}
\section{Dataset and Temporal Network Construction}
\label{sec:dataset}
In this work, we use the Weibo dataset provided by WISE 2012 Challenge\footnote{http://www.wise2012.cs.ucy.ac.cy/challenge.html}. The dataset contains user data and the reposting information of each microblog along with the reposting times which enables us to form the cascade tree networks for each microblog separately.  From the set of 279,148 cascades which spanned between June 1, 2011 and August 31, 2011, we only work with cascades with more than 300 nodes. This leaves us with 7407 cascades.\\
\noindent \textbf{Social Network:} We denote the social network information by an undirected network $G_D$=$(V_D$, $E_D)$ where $V_D$ denotes the individuals involved in the historical diffusion process and an edge $e \in E_D$ denotes that information has been shared between a pair of individuals ignoring the direction of propagation. The diffusion network  $G_D$=$(V_D$ , $E_D)$ is created by linking any two users who are involved in a microblog reposting action within the period May 1, 2011 and July 31, 2011.   We constructed the social network from the historical diffusion network which spanned between June 1, 2011 and August 31, 2011 prior to our cascades in study. It has 6,470,135 nodes and 58,308,645 edges where the average node degree is 18.02.\\
\noindent \textbf{Cascade Network:} A \textit{cascade network} (we refer to these as our diffusion networks) produced by a set of individuals participating in a set of reshares is denoted by  $G^{\tau}_C$ = $(V^{\tau}_C, E^{\tau}_C)$, $V^{\tau}_C$ denotes all the individuals who participated in the diffusion spread of $C$ in its entire lifecycle spanned by $\tau$. An edge $e=(u, v) \in$ $E^{\tau}_C$ denotes that either $v$ reshared $C$ from $u$ indicated by the presence of $(u, v, t) \in \tau$ at some time $t$ or the interaction happened in the past denoted by the presence of $e$ $\in$ $E_D$, that is to say we add the influence of the propagation links from our social network to the current diffusion network - this is done to gauge information diffusion from the topological perspective in the presence of past social network or historical diffusion knowledge. \\
\begin{figure}[H]
	\centering
	\includegraphics[width=12.5cm, height=3cm]{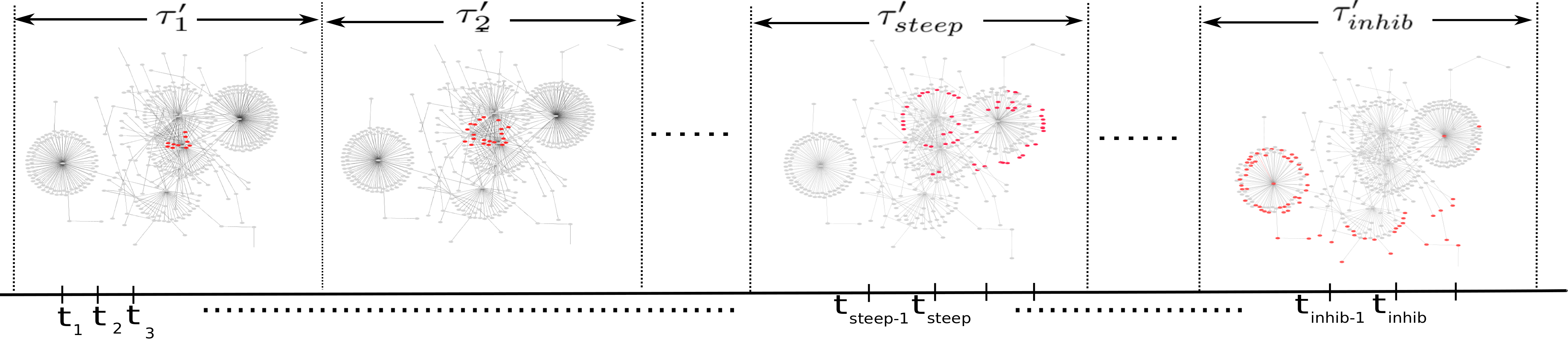}
	\caption{Example showing how the network is partitioned over subsequences for network analysis. Each $\tau'$ contains equal number of nodes for analysis and as the cascade progresses different nodes get activated marked in red. Although each $\tau'$ looks uniform in the time span, we note that each $\tau'$ may differ in the time range depending on how much time it took for $|V^{\tau'}|$ nodes to form the network.}
	\label{fig:net_evolve}
\end{figure}
\vspace*{-6mm}
\noindent \textbf{Temporal cascade network:} In context of $G^{\tau}$,  we denote $G^{\tau'}$ = $(V^{\tau'}, E^{\tau'})$ as the subgraph of $G^{\tau}$, where $V^{\tau'}$ denotes the individuals who reshared a cascade $C$ in the time subsequence $\tau'$ and  $E^{\tau'}$ denotes the set of edges $e=(u, v)$ where the resharing from $u$ to $v$ happened in the time subsequence $\tau'$ or there was an interaction in the historical diffusion period indicated by the presence of $e$ in $E_D$.  For the subsequences, the following conditions hold: (1) $|V^{\tau'_i}| = |V^{\tau'_j}|$ and (2) $E^{\tau'_i} \cap E^{\tau'_{j}}$ = $\emptyset$, $\forall i \neq j, \in [0, \mathcal{Q}]$. We note that the condition $|\tau'_i|$ $\neq |\tau'_j|$ may or may not hold for any $i \neq j$, that is to say the time range spanned by the subsequences in itself may differ depending on the time taken by $G^{\tau'}$ to form the network. In our work, we select and keep $|V^{\tau'}|$ fixed for every cascade in our corpus and set it to 80 nodes. Since we analyze the subsequences in a sequence that depicts the evolution of the cascade over time as shown in Figure~\ref{fig:net_evolve}, the advantage of selecting this subsequence node set size a-priori is that we can avoid retrospective observation of the entire cascade lifecycle and incrementally progress with the network analysis using motifs over intervals until we reach $\tau'_{inhib}$. Since we do not fix $\mathcal{Q}_C$, the number of subsequences for $C$ and let it vary based on $|V^{\tau'}_C|$, we do not fix a specific index $inhib$ $\in [0, \mathcal{Q}]$ for $\tau'_{inhib}$  for all cascades.
\vspace*{-3mm}

\section{Network Motifs} \label{sec:motifs}
Network motifs are recurring induced subgraphs in complex networks that appear significantly higher in count than those in random networks (significance here being determined by $z$-scores with the null model being motifs extracted from a random network with the same degree sequence as the original network, as described in \cite{Milo}). In this paper, for each subsequence network $\tau'$, we extract undirected graph motifs of node size $k=5$. We choose 5-node motifs over smaller sizes since the number of possible patterns occurring from their permutations are too low to conclude anything significant. One of the motivations behind using undirected network motifs in this work is that while direction of information flow is definitely one way to understand the cascade progress, we would like to investigate how users form these micro-scale patterns that constitute the bigger network in these stages. Hence, understanding these motif occurrences helps us understand the overall structural organization of the cascades in terms of the spreading process. We use the FANMOD algorithm because it can detect network motifs up to a size of eight vertices using a novel algorithm called RAND-ESU~\cite{csardi}. We emphasize that the process of extracting the motif patterns is not to count their frequencies - we perform this step to just find the possible patterns for size $k$ motifs. We can substitute it by enumerating motifs in our own way and just finding the instances of such a pattern in $G^{\tau'}$.

\vspace*{-3mm}
\section{Motif Network Coverage}
\label{sec:perc_algo}
Network motifs have been used in understanding the wiring patterns of networks and are known to be basic building blocks of the network formation process. We use network motifs as building blocks to understand how these patterns can build up the network themselves and how close that self-built network is to the original network structure of $G^{{\tau'}_{steep}}$ and $G^{{\tau'}_{inhib}}$. We devise an algorithm to see the spread of these motifs when aggregated over a rolling template in the steep and inhibition intervals.
 \begin{figure}[!t]
	\centering
	\includegraphics[width=13.5cm, height=4cm]{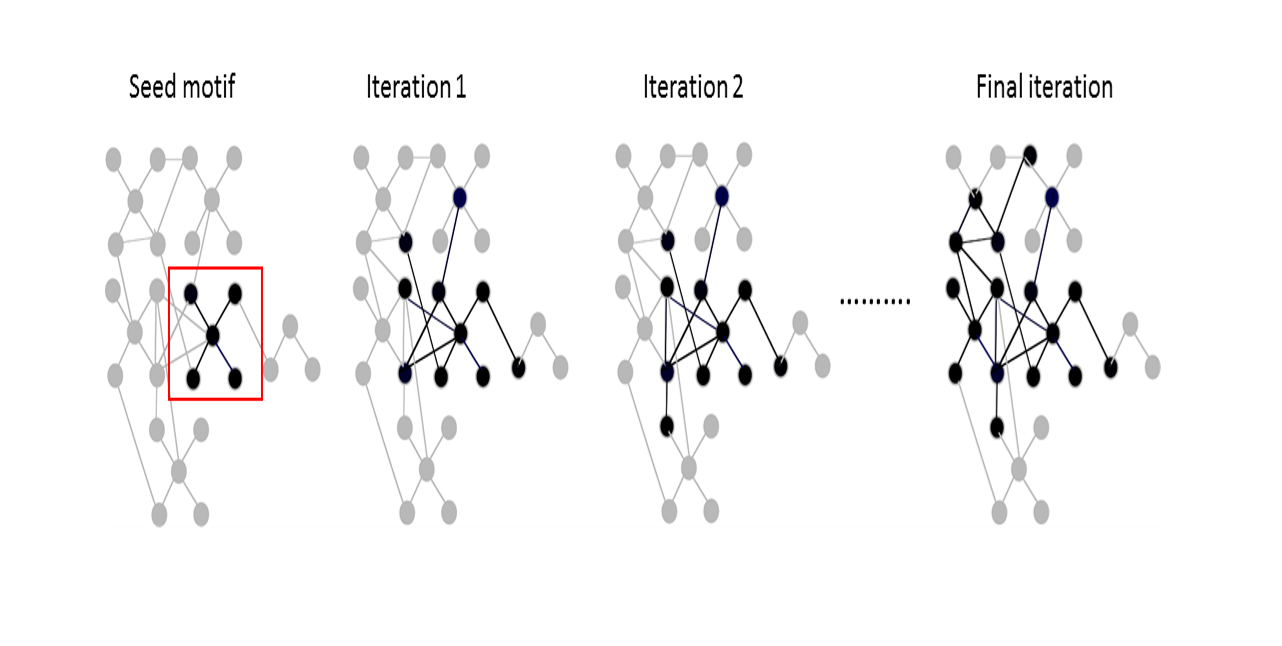}
	\caption{Example  showing how the motif percolation creates a chain of adjacent motifs to form a compact structure.  For each pattern $M$ $\in \mathcal{M}$ that is found in $G^{\tau'}$, we only consider the motif instances $\{ M \}$ belonging to this family $M$- here \  \protect\includegraphics[scale=0.04]{M0.png}.  We start with a seed motif instance shown in the initial configuration of $G^{\tau'}$ and then over each iteration cover instances which are adjacent to the already covered motifs subject to constraints. Once there is no adjacent instance that can be included, the process stops and the covered nodes and edges form $G_{*, M}^{\tau'}$ for that cascade.}
	\label{fig:motif_perc}
\end{figure}
% Algorithm
\begin{algorithm}[t!]
	\KwIn{Motif pattern $M$, $G^{\tau'}$ = $(V, E)$, $\{M\}$ the set of motif instances with pattern $M$.}
	\KwOut{The edges $E_{*, M}^{\tau'}$ $\in E$ that are covered by $\{M\}$ using the algorithm.}
	
	$m_{seed}$ = instance $m$ $\in \{M\}$ having the highest vertex degree sum\;
	$Edges\_covered$ = $E_{m_{seed}}$ \tcp*{edges belonging to $m_{seed}$}
	$Vertices\_covered$ = $V_{m_{seed}}$ \tcp*{vertices belonging to $m_{seed}$}

	\tcc{iterate over all the motifs as long as new edges are added in new iteration}
	\SetKwRepeat{Do}{do}{while}
	\Do{no new edges have been added to $Edges\_covered$}{
		$new\_edges$ = $\phi$\;
		$new\_vertices$ = $\phi$\;
		\For{each motif instance $m$ $\in$ $\{M\}$}{
			$k\_cov$ = number of vertices of $m$ present in $Vertices\_covered$\;
			\tcc{Check if the number of vertices in this new motif instance shared with $Vertices\_covered$ = k-1}
			\If{$k\_cov$ equals $k-1$}{
				$new\_edges$ = $new\_edges$ $\cup$ $E_m$\;
				$new\_vertices$ = $new\_vertices$ $\cup$ $V_m$\;
			}
		}
		$Edges\_covered$ = $Edges\_covered$ $\cup$ $new\_edges$\;
		$Vertices\_covered$ = $Vertices\_covered$ $\cup$ $new\_vertices$\;
	}
	return $Edges\_covered$
	\caption{Motif Percolation Algorithm}
	\label{alg:MPA}
\end{algorithm}
Let $\mathcal{M}_{C}^{\tau'}$ denotes the set of motif patterns that can be extracted using the motif computation algorithm in a subsequence and $M_{C}^{\tau'}$ be one such motif pattern in the set $\mathcal{M}_{C}^{\tau'}$. Let the motif instances for this pattern be $\{M_{C}\}$. Since we evaluate each pattern individually, we will drop the subscripts as usual when generalizing the mechanism. We use an algorithm similar to the Clique Percolation Algorithm (CPM) \cite{der2005} that starts at the maximal clique instance and forms a community of nodes through rolling chains of adjacent cliques. However, since motif patterns do not exhibit a clique structure except for the fully connected motif pattern, we modify this algorithm to start with a seed instance $m_{seed} \in \{M\}$ and then roll the pattern template $M$ through adjacent instances to form a compact network structure in a bottom-up mechanism which we describe in detail next. Let $G_{*, M}^{\tau'} $ = ($V_{*, M}^{\tau'}, E_{*, M}^{\tau'}$) be this structure induced by the set of motifs of pattern $M$ in $\tau'$. To measure the extent to which this induced network is similar to the original network structure $G^{\tau'}$, we define network coverage $NC$ for this motif pattern as $NC_M = \frac{|E_{*, M}^{\tau'}|}{|E^{\tau'}|}$, where both these networks are defined during the same $\tau'$, the numerator represents the edge set cardinality of $G^{\tau'}_{*, M}$ and the denominator refers to that of $G^{\tau'}$. Intuitively, we try to see to what extent does this network generation process based on motifs belonging to each $M$ replicate the original network structure in a subsequence using just one aspect of the network structure, the edges covered. As a matter of fact, different patterns will end up having a different network structure $G_{*, M}$ since we consider each motif pattern separately and only deal with the motif instances belonging to that family each time. Hence we would have different $NC$ for different motif patterns. 

\noindent \textbf{Motif Percolation Algorithm} \\
Since most motif patterns have low edge densities, we can start with different seed motif instances and then roll the pattern $M$ through adjacent instances to end up with different $G_{*, M}^N$ for a single pattern $M$. In this paper, we start with a seed motif $m_{seed}$  based on the following: $m_{seed}$ = $\sum_{v \in V(m), \ argmax_m} d(v)$, $m \in \{M\}$ and $V(m) \in$ $V^{\tau'}$ denotes the vertices of the motif instance $m$. Although we might end up with different components with different seed motifs, we use an iteration algorithm that uses one heuristic that we found better than random motif seeds. The motif percolation algorithm proceeds in the following way: for a pattern $M$ $\in \mathcal{M}$ with node size $k$, two instances $m_1$, $m_2$ $\in \{M\}$ are said to be adjacent if they share $k-1$ nodes and the final motif structure corresponds to the set of nodes and edges in which all the motif instances can reach each other through chains of such adjacent instances starting with $m_{seed}$. A brief description of the algorithm is given in Algorithm~\ref{alg:MPA} and demonstrated in Figure~\ref{fig:motif_perc}. In case there are multiple motif instances with same maximum degree, we pick one of the instance  randomly at the beginning of the procedure. Since there is no way to ensure whether the set of edges selected by the algorithm is maximal, one way to resolve it would to be repeat the entire procedure with different starting seed instances randomly or based on a heuristic. We have made our code publicly available for download\footnote{%\url{https://github.com/SOUMAJYOTI/Information\_Flow\_Motifs}
\url{https://bit.ly/2Tcgae4}}.
\vspace*{-4mm}
\section{Results}
\begin{figure}[t!]
	\centering
	\begin{tabular}{@{}c@{}}
		\includegraphics[width=7cm, height=4cm]{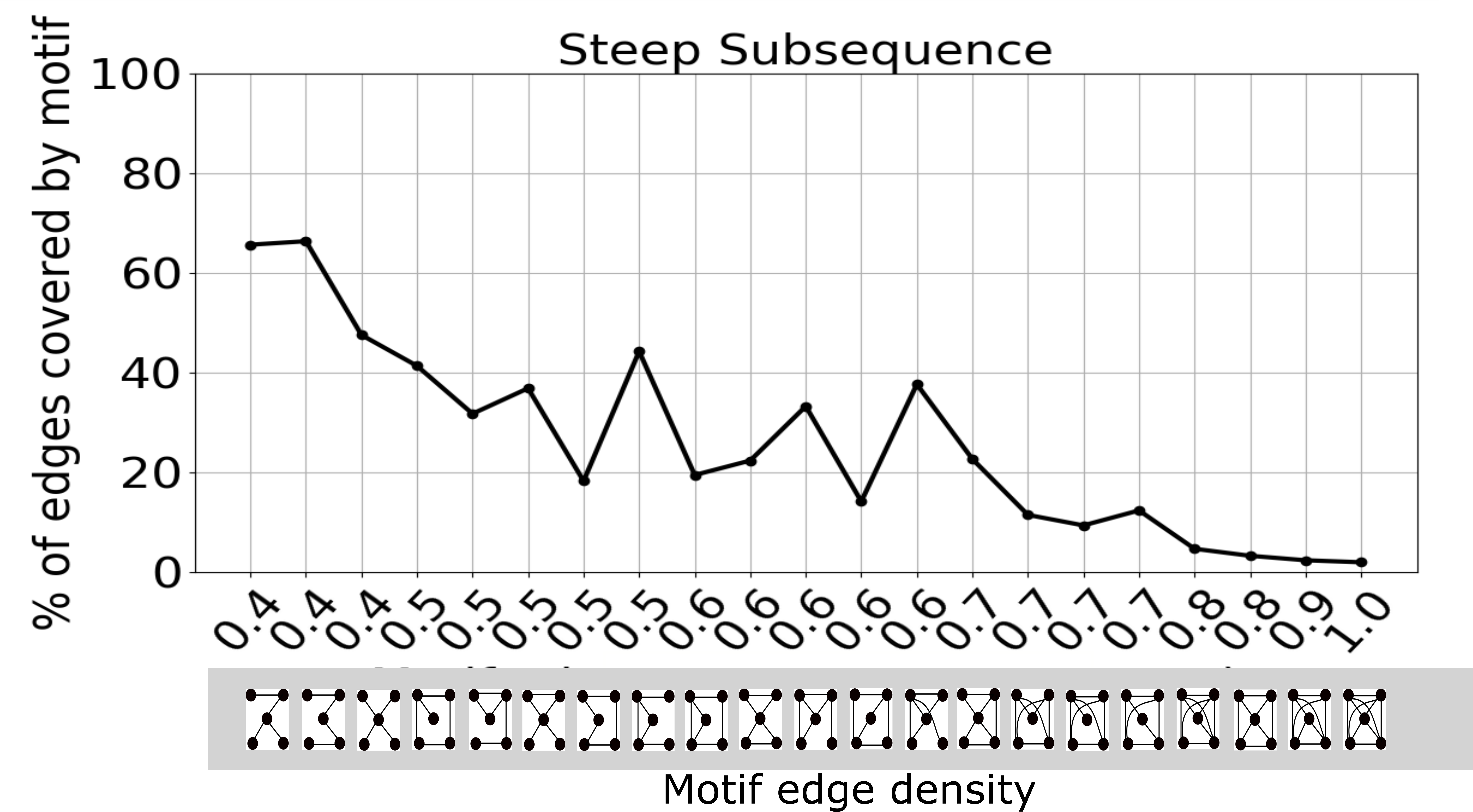} \\[\abovecaptionskip]
		\small (a) 
	\end{tabular}
	
	\vspace{\floatsep}
	
	\begin{tabular}{@{}c@{}}
		\includegraphics[width=7cm, height=4cm]{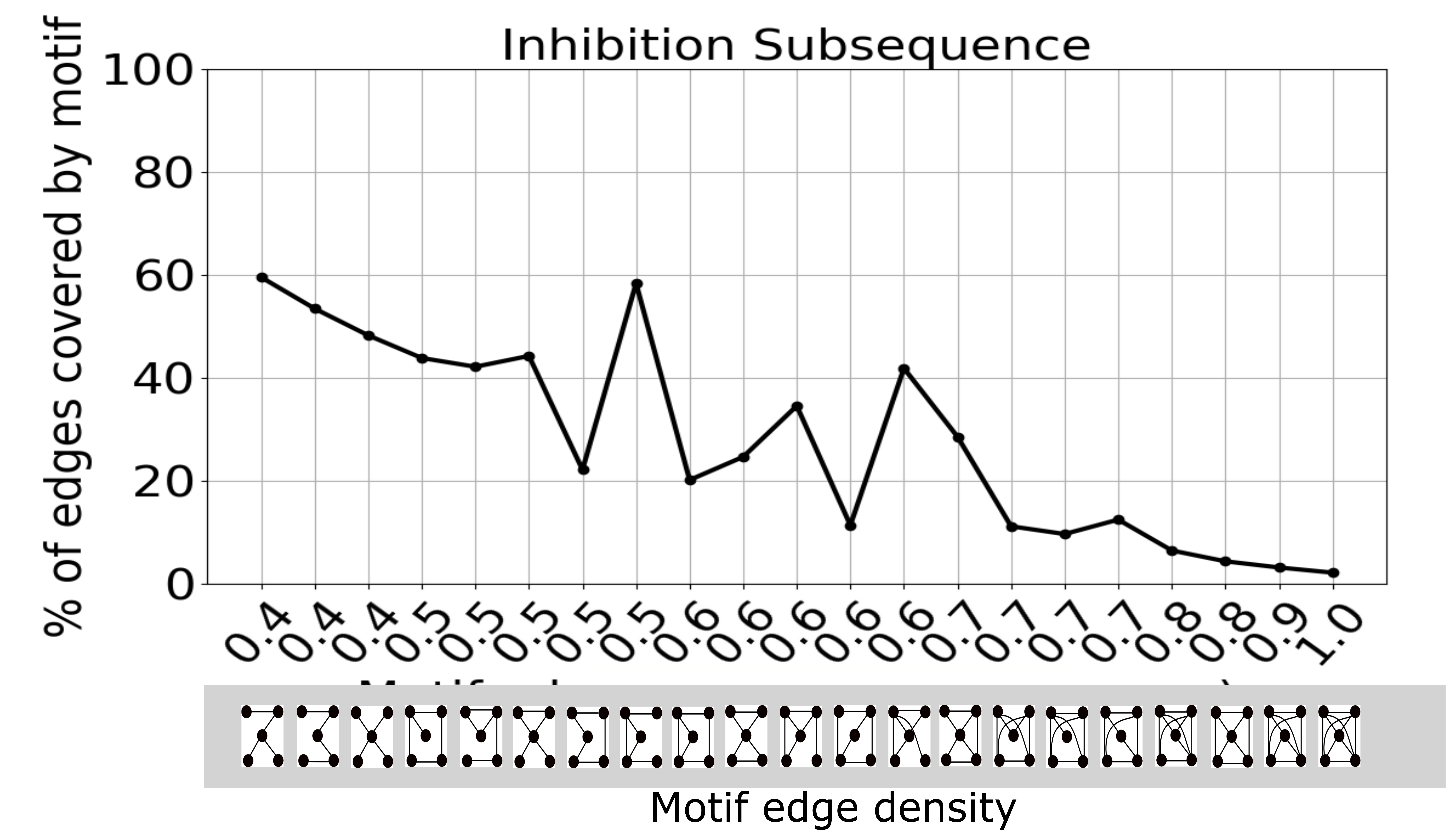} \\[\abovecaptionskip]
		\small (b) 
	\end{tabular}
	
	\caption{(a) The (mean) percentage of edges covered by the percolation algorithm in $\tau'_{steep}$, (b) the (mean) percentage of edges covered by the  algorithm in $\tau'_{inhib}$.}
		\label{fig:acc}
\end{figure}

Before we discuss our results on the motif coverage procedure comparing the diffusion process in the steep and inhibition intervals, we want to emphasize that counting the appearance of subgraph patterns in these subsequences or measuring the graph densities in $\tau'$ does not solve our problem of understanding the network formation process. 
The network coverage on edges and motifs shown in Figure~\ref{fig:acc} presents some key observations with regard to motifs as building blocks: as the density of the motif patterns increases, the network coverage starts decreasing during both the steep and inhibition phases. Therefore, from the perspective of the lowest and largest densities, we do not observe any statistically significant differences. In order to examine the difference between the coverage for the population of cascades with respect to each pattern, we perform a two-sample $t$-test considering the following hypotheses for each pattern M: $\mathbf{H_0}: NC_{M, {\tau'}_{steep}} = NC_{M, {\tau'}_{inhib}}, \ \ \mathbf{H_1}: NC_{M, {\tau'}_{steep}} \neq NC_{M, \tau'_{inhib} }$. Here, we consider the significance level $\alpha$=0.01. The null hypothesis states that the mean network coverage $NC$ for a pattern $M$ is equal for the steep and inhibition subsequences. When we run the tests for each pattern spearately, we observe that surprisingly, the closed loop patterns with high densities and the acyclic patterns with the lowest densities do not have any statistical difference with respect to the difference in coverage values denoted by the fact that the null hypothesis is rejected for them. From Table~\ref{tab:p_val}, we find that only 4 patterns have statistically significant difference when compared: among the cyclic patterns, while for \  \includegraphics[scale=0.04]{M1.png} \ and  \  \includegraphics[scale=0.04]{M0.png} \ which had the highest motif counts (in terms of frequency of appearance) in the both subsequences, the differences were not significant, we find that  \  \includegraphics[scale=0.04]{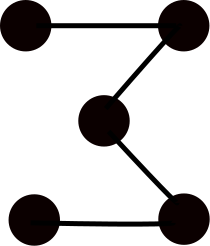} \  exhibits a significant difference in the coverage, with an average of 66.3\% edges covered for all the cascades in $\tau'_{steep}$ and 53.4\% for $\tau'_{inhib}$ - this somehow reiterates the fact the part of structural virality comes from trees having higher depth through linear chains and information flows faster under the exhibition of these linear chains.

On the other hand, among the patterns with loop, we find that some of the patterns with densities on the higher end does not show any difference in the  values exhibited. On the other hand, we find that for patterns  \  \includegraphics[scale=0.04]{M3.png} \,  \  \includegraphics[scale=0.04]{M13.png} \  and  \  \includegraphics[scale=0.04]{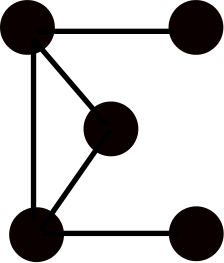} \, the coverage differed significantly with higher coverage exhibited by the inhibition subsequence. This leads to two main conclusions - (1) when the cascade leaves the phase of steep growth and starts approaching the inhibition phase, the network organization is represented through communities that are held by these motif patterns more. So the diffusion mechanism during the approaching inhibition phase is characterized by the information adopted by  users who are connected to other users through these patterns, and (2) note that since loops indicate the presence of social network links  in a motif pattern (in the absence of social network links, cascades exhibit a default tree structure), it reinforces the observation that during the later stages of the cascade lifecycle, users start adopting when there have multiple friends or followees who have adopted the cascade and thus multiple exposures needed for information flow slows the growth. 
%\vspace*{-5mm} 
\begin{table}[!h]
	\centering
	\begin{tabular}{|l|l|l|l|l|l|l|l|l|l|l|l|}
		\hline
		& \multicolumn{11}{c|}{\textbf{Motif Patterns}}                                                                        \\ \hline
		& \includegraphics[scale=0.06]{M1.png}    & \includegraphics[scale=0.06]{M2.png}    & \includegraphics[scale=0.06]{M0.png}              & \includegraphics[scale=0.06]{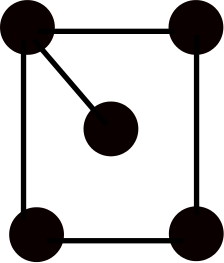}     & \includegraphics[scale=0.06]{M13.png}              & \includegraphics[scale=0.06]{M3.png}              & \includegraphics[scale=0.06]{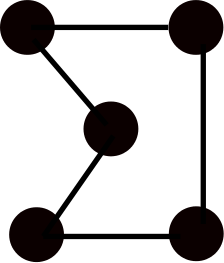}    & \includegraphics[scale=0.06]{M15.png}              & \includegraphics[scale=0.06]{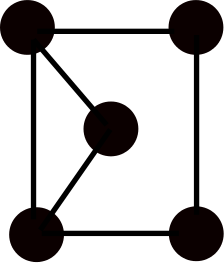}    & \includegraphics[scale=0.06]{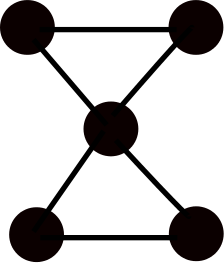}   & \includegraphics[scale=0.06]{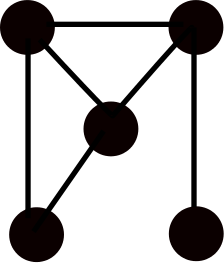}    \\ \hline
		\textbf{p-values} & 0.18 & \textbf{0.071} & 0.06 & 0.097 & \textbf{0.008} & \textbf{0.007} & 0.13 & \textbf{0.004} & 0.11 & 0.18 & 0.094 \\ \hline
		& \includegraphics[scale=0.06]{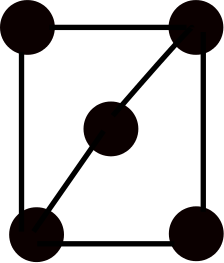}   & \includegraphics[scale=0.06]{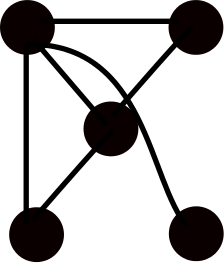}    & \includegraphics[scale=0.06]{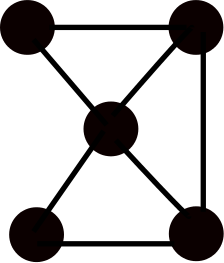}             & \includegraphics[scale=0.06]{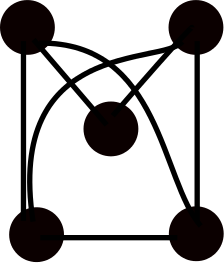}    & \includegraphics[scale=0.06]{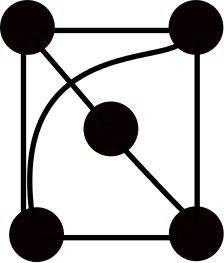}             & \includegraphics[scale=0.06]{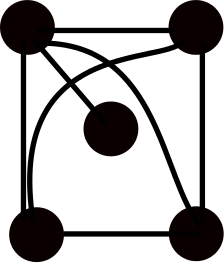}             & \includegraphics[scale=0.06]{M10.png}   & \includegraphics[scale=0.06]{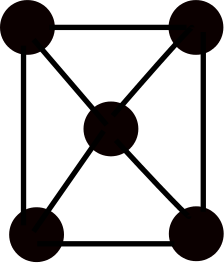}             & \includegraphics[scale=0.06]{M14.png}   & \includegraphics[scale=0.06]{M16.png}   &       \\ \hline
		\textbf{p-values} & 0.07 & 0.16 & 0.18           & 0.2   & 0.367          & 0.293          & 0.21 & 0.30           & 0.25 & 0.17 &       \\ \hline
	\end{tabular}
\caption{$p$-values for the two-sample t-test that measures the difference in the NC values for the respective patterns between the steep and the inhibition subsequence.}
\label{tab:p_val}
\end{table}
%\vspace*{-1mm} 
To recap, these results suggest that the diffusion mechanism in these two stages are differentiated not by the presence of dense patterns that blocks information flow outside their communities nor by just presence of star patterns signifying one central user in the steep phase, but they are distinguished by some specific patterns. It hints at the conclusion that the diffusion process in the inhibition phase is modulated by mainly triads with small groups who choose to share information amongst one another.
\vspace*{-4mm}
\section{Related Work}
\vspace*{-3mm}
Our work is closely related to the field of community detection which enjoys a prolific literature. In particular, community detection in static networks has received much attention from multiple disciplines including sociology and computer science. This has resulted in a diverse set of existing methods including traditional network structure-based studies~\cite{Girvan2002}, optimization-based techniques~\cite{Lancichinetti2011,Alvari2011,Chen2010}, label propagation methods~\cite{Raghavan2007,Xie2011} and propinquity~\cite{Zhang2009}.  Also, some other studies have studied the evolution of communities over time and in dynamic networks. The work of~\cite{Hopcroft2004} for example, detected natural communities that were stable to small perturbations of the input data. Later, new communities detected were matched to earlier snapshots using the natural community structure tree. Another work of~\cite{Palla2009} which in essence is related to our work, proposed a method to detect communities in dynamic networks based on the \textit{k}-Clique Percolation Method (CPM). In their approach, communities were defined as adjacent k-cliques that share k-1 nodes. More specifically, two nodes belong to the same community if they can be connected through adjacent k-cliques.
\vspace*{-4mm}
\section{Conclusion}
\vspace*{-3mm}
We leveraged network motifs to investigate whether or not they could explain the diffusion process in cascades. The ability of these patterns to characterize the inhibition stage further leads us to conclude that during the inhibition phase, users need multiple exposures and hence more social reinforcement to adopt an information, thus explaining the slowing down. In contrast, we do not find compelling results for the steep phase of the cascade lifecycle with specific patterns. 

\noindent \textbf{Acknowledgements} Some of the authors are supported through the  ARO grant W911NF-15-1-0282.
\vspace*{-5mm}
\small{
}
\end{document}